\documentclass[12pt,preprint]{aastex}

\usepackage{lineno}

\newcommand{\btxt}[1]{{\bf #1}}

\shorttitle{Pressure Shift and Gravitational Red Shift}
\shortauthors{Halenka et al.}

\begin{document}


\title{
Pressure Shift and Gravitational Red Shift of Balmer Lines in White Dwarfs. Rediscussion
}


\author{Jacek Halenka and Wieslaw Olchawa}
\affil{Institute of Physics, University of Opole,\\ ul. Oleska 48, 45-052, Opole, Poland}
\email{halenka@uni.opole.pl}
\email{wolch@uni.opole.pl}

\author{Jerzy Madej}
\affil{Astronomical Observatory, University of Warsaw,\\ Al. Ujazdowskie 4, 00-478 Warszawa, Poland}
\email{jm@astrouw.edu.pl}

\and
\author{Boleslaw Grabowski}
\affil{Wroclaw School of Information Technology WWSIS  ``Horyzont'', \\ ul. Wejherowska 28, 54-239 Wroclaw, Poland}
\email{bgrab@uni.opole.pl}

\slugcomment{\small \btxt{In Memory of Jan Jerzy Kubikowski (1927--1968) --- one of the Pioneers of plasma in astrophysics}}



\begin{abstract}
The Stark-induced shift and asymmetry, the so-called pressure shift (PS) of H$_{\alpha}$ and H$_{\beta}$ Balmer lines in spectra of DA white dwarfs (WDs), as masking effects in measurements of the gravitational red shift in WDs, have been examined in detail.
The results are compared with our earlier ones from before a quarter of {a century} 
(Grabowski et al. 1987, hereafter ApJ'87; Madej \& Grabowski 1990).
In these earlier papers, as a dominant constituent of the Balmer-line-profiles, the standard, symmetrical Stark line profiles, shifted as the whole by PS-effect, were applied to all spectrally active layers of the  WD atmosphere. At present, in each of the WD layers, the Stark-line-profiles (especially of H$_{\beta}$)
are immanently asymmetrical and shifted due to the effects of strong inhomogeneity of the perturbing fields in plasma. To calculate the Stark line-profiles in successive layers of the WD atmosphere  we  used the modified Full Computer Simulation Method (mFCSM), able to take adequately into account the complexity of local elementary quantum processes in plasma.
In the case of the H$_{\alpha}$ line, the present value of Stark-induced shift of the synthetic H$_{\alpha}$ line-profile
is about twice smaller than {the previous} one (ApJ'87) and it is negligible in comparison with the gravitational red shift. In the case of the H$_{\beta}$ line,
the present value of Stark-induced shift of the synthetic H$_{\beta}$ line-profile is about twice larger than the previous one. The source of this extra shift is the asymmetry of H$_{\beta}$ peaks.
\end{abstract}


\section{Introduction}

Quarter of a century has passed since we got to know for certain that the Stark-induced shift and asymmetry effects (hereafter called the ``pressure shift'', PS), caused by local atomic quantum processes in plasmas, play an important role in formation of the Balmer hydrogen lines in white dwarfs (WDs) spectra, and that contribution of PS to the residual (free of the Doppler component) red shift of these lines must be taken into account while  separating the gravitational (Einstein's) red shift. We meet the first note on that theme in the papers by \citet{wie71} and \citet{wie72}. They experimentally revealed systematic slight shifts of these Stark-broadened lines to the red, which are linear functions of the electron density. As the source of light they used hydrogen plasma --- the environment similar to that in the white-dwarf atmospheres. Moreover, the measurements 
were taken in the same way as those in the WD spectra in two pioneering papers by \citet{gre67}, and \citet{tri72}, where the gravitational red shifts were determined within the relatively broad central region of a Balmer line --- the so-called ``averaged line center'' (ALC) --- in the best spectra  available at that time, i.e. in those recorded by means of a 508-cm telescope.

The  results of the gravitational red shifts determination by Greenstein and Trimble for white dwarfs revealed 
an unexpected and surprising problem, with time called \textit{the astrophysical puzzle} \citep[cf.][]{wei79}: due to some unknown reasons, the measured (residual, i.e. free from the Doppler component) WD red shifts were systematically too great, up to about 50 per cent (10 -- 15 km/s in Doppler velocity units) exceeding predictions by Einstein's General Relativity (GR). (More precisely: WD masses resulting from the gravitational red shift measurements systematically exceeded those determined using the so-called astrophysical methods, i.e. astrometric one, model atmospheres, and so on.)  \cite{wie71}, in their paper {under} the significant tittle ``On the cause of the redshifts in white-dwarf spectra'', noted \textit{a propos}: ``It appears that the Stark-induced red shift accounts for a portion of the observed red shift in white dwarfs which has previously been attributed entirely to gravitation.''  

At that time, the appearance of the Stark-induced red shift was surprising indeed, because splitting of Balmer lines in external homogeneous electric fields, as observed in those days in  laboratories, was symmetrical.
 Quadratic Stark effect generates line shifts (usually to the red) and line asymmetry. ``However --- we quote \citet{wie71} again --- based on all the then available experimental and theoretical knowledge --- the works of \citet{rau29} and \citet{gri64} are especially cited --- the Stark shifts were estimated to be insignificant.''

In both objects under consideration --- in the WD spectra and in the light of the device (the high-current, wall-stabilized arc) of Wiese-Kelleher's (1971) type --- the hydrogen Balmer lines are enormously broadened by the interparticle Stark effect in plasma. In both cases also these lines, for a long time, appeared to be practically symmetrical and unshifted, contrary to the so-called ``metal'' lines of all heavier atoms --- narrow and easy to measure, but strongly affected by Stark shift and asymmetry. These last (``metal'') lines are completely useless 
for the purpose of the gravitational red shift measurements. For about four decades now it has  undoubtedly been a well-known fact that the Stark-broadened hydrogen Balmer lines in dense 
plasma\footnote{One routinely reads the term ``dense plasma'' (or ``high-density plasma'')  as a plasma of electron concentration about and above $10^{16}$ cm$^{-3}$. For example, in the paper by \citet{wie72}, {entitled} ``Detailed Study of the Stark Broadening of Balmer Lines in a High-Density Plasma'', investigations cover the range of electron densities $N_e$ between $1.5 \times 10^{16}$ and $10^{17}$ cm$^{-3}$. In WD atmospheres, the range of $N_e$ is usually considerably broader.} 
are --- apart from their enormously great broadening --- slightly asymmetrical and red-shifted as well. The asymmetry and red shift are, relatively, of small-size effects, but phenomenally, in their physical specification, very complicated ones. Detailed measurements and a careful theoretical balance of physical processes forming the hydrogen spectral lines in plasma step by step lead us to complete understanding  and to a successful description of the Stark-induced shift. So, the Stark-induced shift, as a masking effect, can be definitely separated from the gravitational red shift measurements in the WD spectra.

The first simple attempt to provide a theoretical explanation of the effect discovered by Wiese et~al.~(1971; 1972) in the hydrogen Balmer lines in plasma was made by \citet{gra75} --- as a specific consequence of  Debye screening, i.e. of the neutralizing action of a negatively charged plasma background on the atomic nucleus, and of resulting  deformation of the internal structure of disturbed hydrogen atoms. The result, PS~$\propto N_e/T$ (for details see that reference), was surprisingly fair: the predicted pressure shift of the line center to the red became evident and was of the same order of magnitude as the observed excess over gravitational red shift. In the recapitulating paper by \cite{gra87}, the synthetic Balmer line-profiles in WD spectra were calculated for a wide range of $\hbox{log}\,g$ and $T_{\hbox{\scriptsize eff}}$ values of WD atmospheres in a physically more realistic approach, corresponding to the world state of the art in the hydrogen-spectral-line broadening theory of that time (see Section~\ref{sec3}). We used, namely, symmetrical Stark-broadening-line-profiles \citep[following][]{edm67} as functions of the local physical conditions in subsequent layers of a stellar atmosphere, in each layer, moreover, shifted as a whole on the value of the pressure shift, PS, dependently on proper local physical condition in the layer. 
The PS-function was calculated, taking into account complex influences of plasma environment on a H atom: (1) predominating effect --- the screening of the atomic nucleus by the cloud of free electrons, following 
\citet{gra75}, but corrected empirically (by a multiplicative factor), accordingly to the laboratory data accessible at that time, and (2) the phenomena of the correcting nature (especially in the case of 
{the H$_{\beta}$ line}), caused by contribution of the resultant quasi-static local ionic field.
{We took into consideration following contributions: multipole expansion of the perturbing fields, the quadrupole term including, and the second order of ionic field's corrections in frequency shift (quadratic Stark effect). The first order correction in the oscillator strength alterations of the line components, and the so-called pressure ionization in plasma, have been taken into account also (for details see \cite{gra87}).}   

The principal results were as follows: Asymmetry and pressure shift (to the red) of the synthetic Balmer line increase with the distance from the line center; {the H$_{\beta}$ line} is strongly affected by PS and, consequently, is less useful for the gravitational red shift determination, especially when far wings are used to define the averaged line center. The relative significance of the pressure shift of Balmer lines, as compared with the Einstein's gravitational red shift, systematically decreases with increasing values of  $T_{\hbox{\scriptsize eff}}$ and  of $\hbox{log} \,g$ --- from about 50 per cent of the gravitational red shift down to about mere 1 percent --- at both opposite poles, i.e. at extremely low and extremely high values, respectively (cf. \citet{mad90}).

Today, over {a quarter} of a century later, when excellent, high-resolution spectroscopic data, with clearly revealed, sharp non-LTE, line centers of Balmer lines are accessible from HST \citep[e.g.][]{bar05}, ESO \citep[e.g.][]{fal10}, or VLT \citep[e.g.][]{cas09},  the importance and usefulness of these old, above-discussed, results on PS are rather archival. At present we are able to calculate the synthetic WD Balmer line-profiles with incomparably higher precision, taking into account a more complete description of the quantum physical processes, active in the WD Balmer line creation. (For comment on the specific phenomena, active in creation of PS and asymmetry of Balmer lines in plasma, taken into account in the present calulations, see Sec. 2, the second through the last paragraphs, especially the visualization in Figs.~\ref{fig1}  and~\ref{fig2}, and also {in} Sec.~\ref{sec3}, the third and the fourth paragraphs.)  

Measurements of the gravitational red shift give WD masses (precisely: mass/radius ratio) immediately and with the highest exactness (of the same order as --- in the  incidental cases ---  the orbit astrometry of binary systems), practically independently of anything --- under condition, however, that the spectral measurements are free (separated from) of disturbing and masking effects, produced by mixed up elementary processes in plasma of WD atmospheres, especially at low values of $\hbox{log}\, g$ and $T_{\hbox{\scriptsize eff}}$. We devote our paper to improving the reliability of this excellent method. Activity in that direction is desired and expected  
\citep[cf. e.g.][]{koe87, pro98, bar05, fal10}. We will discuss relevant problems in next sections.

\section{An introductory note on the physical elementary processes forming the
Balmer line profiles in plasma (in WD atmospheres in particular)}

One cubic centimeter in a sample layer of the typical WD atmosphere, at temperature of the order of  ten thousand Kelvin, 
contains about $10^{20}$ atoms, almost all are hydrogen ones, of which one in a thousand is an ionized atom, i.e. without a bound electron.  So, in that volume we meet about $10^{17}$ electric charges (electrons and ions). The electric field, attached to a charged particle, is a long range one ($\sim r^{-1}$), therefore each charged particle acts simultaneously on lots of other charges and neutrals. All particles run with great velocities, of hundreds/thousands of meters per a second. In the scale of nanoseconds (the typical clock beat for atomic micro processes), the influences by ions (slow), are nearly constant,  resulting in quasi-static splitting of the atomic levels into sets of sublevels; the perturbations by electrons (fast) --- rapidly changing, are ``seen'' by atoms as impacts, resulting in a reduction of life-times and in broadening of the atomic sublevels. Moreover, mutually overlapping field strengths reduce the scope of this field penetration (the so-called screening effects). This is a typical scenario of line profile formation in the plasma conditions.

\begin{figure}
\plotone{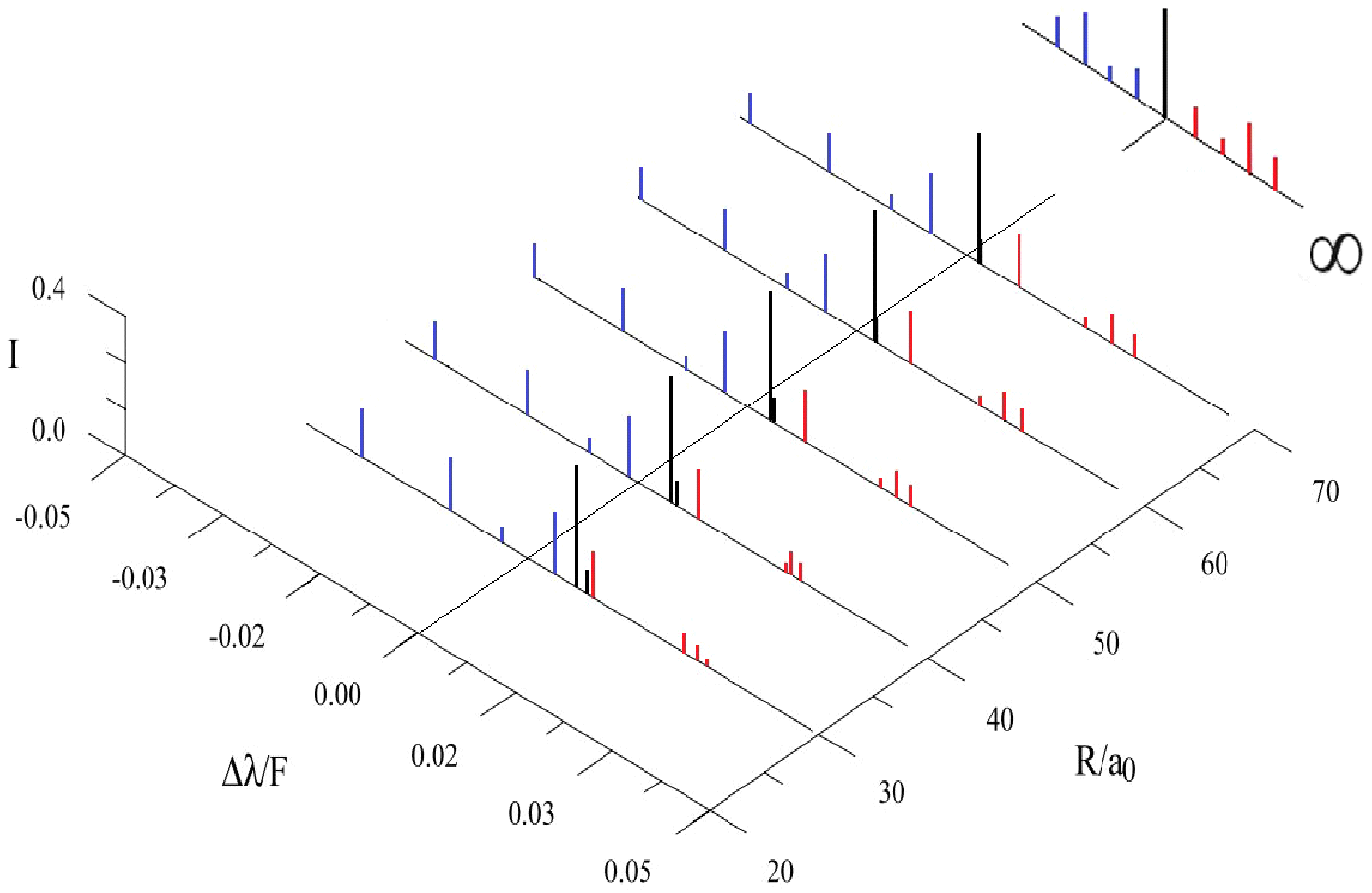}
\caption{The quasi-static Stark spectrogram (pattern of splitting) of the H$_{\alpha}$ Balmer line, as an example, in the field of a single, fixed ion, as a function  of the ion-atom distance $R$, expressed in the reduced scale $R/a_0$, where $a_0$ is the Bohr radius; at (formally) $R/a_0 = \infty$ (in the homogeneous perturbing electric field) --- the canonical picture of the linear Stark effect (LSE) splitting. The distance from the line center is expressed in a reduced scale (in the denominator
there is the field strength in the corresponding ion-atom distance).
The blue and red components are marked by proper colors, the central component –-- by black.
\label{fig1}}
\end{figure}

\begin{figure}
\plotone{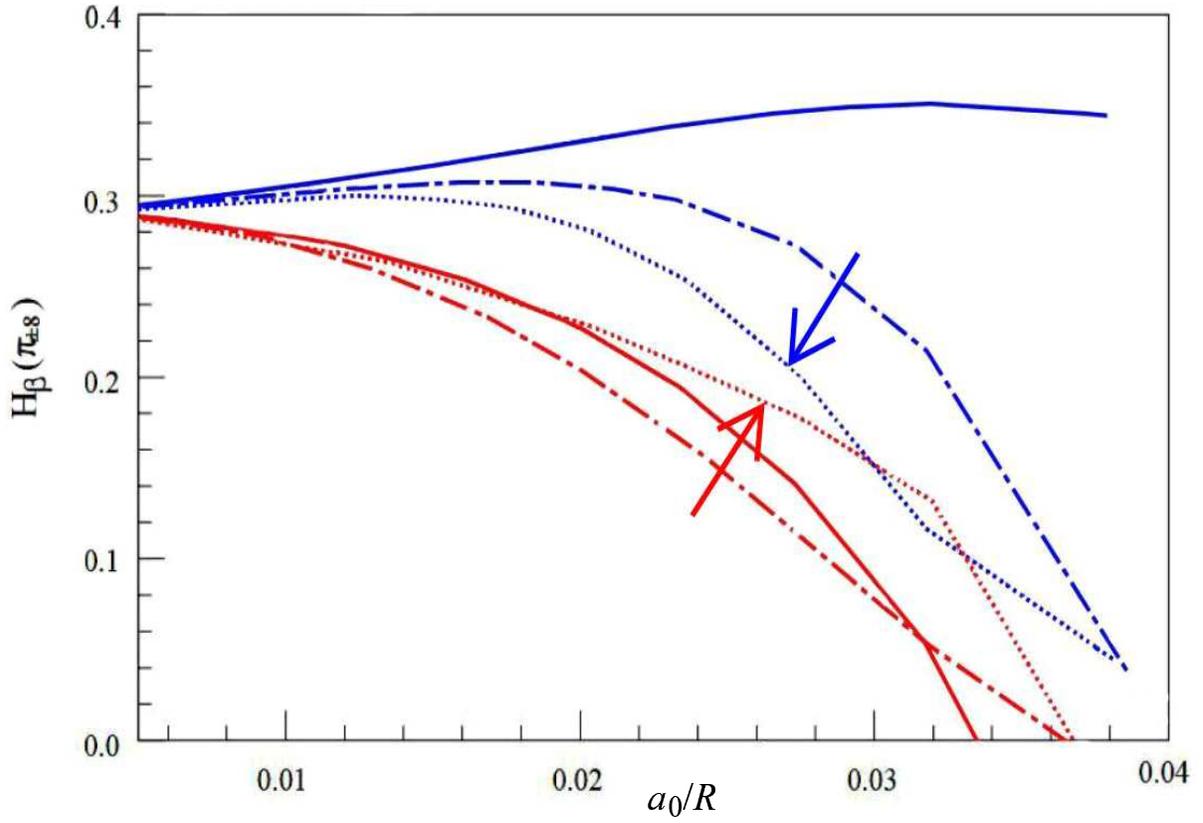}
\caption{Evolution of the intensity of selected Stark components of the H$_{\beta}$ line as a function of the reciprocal ion-atom distance in different approximations. For details --- see text.
The blue and red components are marked by proper colors.
\label{fig2}}
\end{figure}

For start, we consider the simplest, virtual case: a single ion (proton) at rest, acting on a tentative hydrogen atom, also fixed, dependently on the distance between these particles. We temporarily ignore all the remaining charged particles. This pair of particles (called the nearest neighbor approximation, NNA) stands for an extremely simplified physical quantum system,  but the mathematical problem of solution of the corresponding time-independent perturbation theory is by no means simple \citep[cf.][and Figs.~\ref{fig1} and~\ref{fig2} therein, and below]{olc01}.  

Fig.~\ref{fig1} presents the solution for {the Balmer-$\alpha$ line} as an example. In the homogeneous external electric field (in our demonstrative model of the perturbing point charge, the homogeneous field formally corresponds to the infinite distance of the perturber) the splitting of line components is strictly {symmetrical} (see up-plane of the figure). This is the canonical pattern of the linear Stark effect (LSE) of Balmer-${\alpha}$ in a homogeneous external field.
The field of the perturbing ion is, of course, the more inhomogeneous, the nearer the ion resides (the ion-atom distance is expressed in $R/a_0$, $a_0$ --- Bohr radius) and, consequently, the multipole expansion (ME, some kind of the Taylor series) of the perturbing field strength must be taken into account in the calculations, i.e. the dipole term (corresponding to the homogeneous component), a quadrupole one (the first order correction), an octupole one (the second order correction), and so on, if necessary. In the case under examination, all these higher order multipole terms have been taken into account. We see the effect of these terms on the line splitting
in the changes in location (and even in succession!) and in intensities of the line components. Particularly, the intensities of the red components (the right hand side of the {figure}, when we are looking toward the LSE pattern) decrease with the decreasing ion-atom distance, but all line components move to the red --- creating a complicated image of the pressure asymmetry and shift formation.

In Fig.~\ref{fig2},  the evolution of the intensities of the selected Stark components of the H$_{\beta}$ line is presented demonstratively. As a sample, the strongest ($\pi_{8}$)  Stark components are used: $\pi_{-8}$ (the blue component, the upper curve of the given mark in the figure), and  $\pi_{+8}$ (the red component, the lower curve in the figure), in dependence on the approximation used, as a function of the reciprocal reduced distance ($a_0/R$). The marks used have the following meaning: solid lines --- dipolar perturbation (homogeneous external field), the quadratic Stark effect is included; the short-dashed curves --- the quadrupole interaction 
is also included; the dotted curves --- the octupole interaction is also included. The evolution presented by the dotted curves (marked by proper arrows) is the most precise.

In Fig.~\ref{fig1} and, especially, Fig.~\ref{fig2}, we see that the main source of asymmetry --- and of the asymmetry-induced shift --- in the quasi-static approximation is the inhomogeneity of the ion microfield in {the surroundings} of an atom, active in the spectral line formation: the quadrupole and the octupole terms, and also the quadratic Stark effect.

\section{Calculations of the Stark profiles of hydrogen lines in plasma environment}
\label{sec3}
The fundamental principles of the Stark-profile calculations of hydrogen lines, formed in the plasma environments, were formulated in the 1950s, in the pioneering papers by Michel \citet{bar58} and in the 1960s by Hans R. Griem and by his coworkers \citep{gri62}. In the so-called generalized impact-broadening theory, formulated at that time, the influences of the fast moving electrons perturb, as impacts, a hydrogen atom and reduce life-time (causing broadening) of sublevels (separated and well-defined in quasi-static ion field) of both the upper and the lower levels of the line. A complete Stark profile is a result of an averaging of impact profiles created in such a manner over the quasi-static ion field strengths within the Debye's sphere (some ``interacting volume''  of plasma round an atom, limited by effects of mutual screening of the charged particles). The best sources of the Stark profiles of those days were these by \citet{gri64}, \citet{gri74}, and \citet{kep68}, generally known as ``modified impact theory''. Nearly equivalent ones are: the ``unified theory'' by \cite{vid73}, and ``semiempirical profiles'' by \citet{edm67}. They account for the Debye's shielding  effects of the perturbing ions, but the ions everywhere are treated as static ones (the thermic ion dynamics is omitted generally), and, furthermore, the Stark splitting caused by ions is assumed as LSE entirely --- a symmetrical and unshifted one ---  i.e. the higher order corrections in the multipole expansion, ME, and in the perturbation theory, PT (the {second} order PT-correction gives the quadratic Stark effect, QSE), are completely neglected.
In all these calculations, the classical path approximation, CPA (assuming that the perturbing electrons move along the classically specified ways) is always used.  Fortunately, CPA is entirely satisfactory in most applications \citep[e.g.][]{gri74} and does not demand improvement. Negligence of the ion dynamics creates, however, serious uncertainty in the line centers --- in the region of special importance for gravitational red shift measurements. An extreme defect --- from the point of view of the gravitational red shift measurements in WD spectra --- of the quasi-static approaches in the majority of existing Stark-profile theories lies in the complete negligence of the higher order corrections in the Hamiltonian (ME) and in PT, as well. It is these corrections that produce the red shift and asymmetry features in the hydrogen-line profiles, observed in spectra of moderately dense and in high-density plasmas, and, naturally, appearing in the WD spectra, too. 

The spectral line shapes, particularly the hydrogen line shapes, are one of the most important carriers of information on the structure of laboratory, as well as on space plasmas. From the astrophysical point of view, it refers mainly to the plasma of the stellar atmospheres and the plasma of WD atmospheres. In order to get reliable recognition of the physical features of the investigated plasma environment (i.e. in order to get the plasma diagnostics) through the line-shape analysis, the line shapes must be calculated with sufficient accuracy. To attain this  different approaches are accepted --- analytical methods and computer codes of various complexity and accuracy, and of different limits of applicability. The reader, interested in this subject, can find more information on the line broadening  in recent reviews, e.g. by \citet{gue97}, \citet{dju09} and in papers quoted therein. Particularly,
one can find a review of the line profiles obtained by using the so-called Full Computer Simulation Method (FCMS) compared with other approaches in the paper by Ferri et al. (2014). 

The line shape and/or the line width (full-width-at-half-maximum, FWHM, in particular) serves the purpose of  plasma diagnostics, as the main informer on the physical properties of a diagnosed medium. FCSM
approaches \citep{gig89,gig96,hal96}  
very faithfully reconstruct the local phenomena in plasma and, consequently, give shape and width of the line relevant to this aim. These approaches are, however, able to predict neither asymmetry nor a shift of the line. 

We refer shortly to the principle of FCSMs. The local  ionic, as well as electronic, shares in the spectral line formation are appointed using the simulation method, and the time dependent Schr{\"o}dinger equation is solved numerically. The ionic contribution is, as a rule, limited to dipole interaction of the atom-ionic field, and to the first order correction in PT, i.e. to the linear Stark effect (LSE) only. The resulting line-profile is therefore a symmetrical and unshifted one. However, the FCSMs have {no limitations} present, e.g., in \citet{kep68}, and \citet{vid73} theories, i.e. quasi-static ion fields, nor an impact approximation for free electrons. 
The quadrupole interactions (the first order correction in the multipole expansion, ME), as well as the quadratic Stark effect (QSE, the second order correction in PT) can be  taken into account in FCSM, indeed --- see paper by {\citet{olc02}}. Such an improved approach leads to effects of the line-asymmetry and to Stark-induced shift --  to effects important especially in conditions of dense plasma. These predictions excellently agree with the recent laboratory measurements. Unfortunately, this improved approach is very time-consuming, and a laborious one. We, therefore, have {developed} an equally reliable, but more effective method, useful in efficient line-profile calculations, when a wide range of different physical conditions (successive layers of the WD atmosphere) must be examined. Here the shape of the spectral line is a sum of two components. The first component (of the zeroth order term in PT and the first in ME, i.e. LSE in dipole approximation), is calculated using 
{the FCSM} approach. The second component {(consisting of a variety of corrections, each small compared to the first component)} includes the contributions originating from higher-order terms of the emitter-ionic microfield interactions (the quadrupole interactions and QSE, calculated in quasi-static approximation for ions). This second component describes the shift and asymmetry of the Balmer line profile, produced by {the ion} constituent of plasma. The shifts caused by electrons have been taken into account, following the paper by \citet{gri83}. The method used —-- modified FCSM (in our notes, mFCSM) --— is an effective one, producing realistic line-profiles: broadened, shifted, and asymmetrical, agreeing, moreover, with the laboratory measurements. This method has been formulated in a series of papers by \citet{hal96}, \citet{olc02}, \citet{wuj02} and \citet{olc04}, and, finally, was
successfully examined and improved in the paper by \citet{gri05}. All the calculations of the hydrogen line-profiles in the present paper have been carried out following this mFCSM method.

Particular problems, related to the matter in the hand, have been solved in the following references: A detailed discussion and proper quadrupole and octupole corrections of the ionic fields, producing asymmetry and red shift of the Stark line profile in plasma --- in a series of papers by Halenka and Olchawa: \citet{hal90}, where correction of the ``modified impact theory'' was made, including the emitter-ion-quadrupole interactions, the quadratic-Stark-effect, and the ion dynamics effects; \citet{hal07} --- octupole inhomogeneity tensor of ionic microfield in Debye plasma at neutral emitter, calculated for the first time; \citet{hal09} --- the octupole inhomogeneity tensor at ionized emitter, also calculated for the first time.

The starting point of this consideration is the relation
between the spectral line profile and the average of the
dipole autocorrelation function $C(t)$, which can be written in
the following way:
\begin{equation}
    I(\Delta\omega)=\lim_{t_{f} \to \infty} \pi^{-1} \int_0^{t_f}C(t)\,\hbox{e}^{\hbox{\scriptsize i}\Delta\omega t} \hbox{d}t
\end{equation}
\begin{equation}
    C(t)=\hbox{Tr}\{\mathbf{d}_{if}\cdot U^\dagger_{ff^\prime}(t)\,\mathbf{d}_{f^\prime i^\prime}\,U_{i^\prime i}(t)\}_{av}/\,\hbox{Tr}\{\mathbf{d}_{if}\cdot\mathbf{d}_{fi}\},
\end{equation}
where $\mathbf{d}$ is the dipole operator for the hydrogen atom, while
$ii^\prime$ and $ff^\prime$ indicate the sublevels of the initial ($E_i$) and final
($E_f$ ) states of the unperturbed atom, respectively. The frequency
separation from the line center is given by $\Delta\omega=\omega-(E_i-E_f)/\hbar$, whereas $U(t)$ is the operator of the time development of the hydrogen atom in the presence of the electric field produced by electrons and ions. The averaging $\{\cdots\}_{av}$ is taken over all initial simulated field strengths and
possible time histories. The time-evolution operators $U_{i^\prime i}(t)$
and $U^\dagger_{ff^\prime}(t)$ (corresponding to the initial and final states, respectively)
satisfy the following Schr{\"o}dinger equation:
\begin{equation}
    i\hbar \dot{U}(t)=[H_0+V(t)]\,U(t),
\end{equation}
where $H_0$ is the Hamiltonian of the isolated radiator and
$V(t)$ is the radiator-plasma interaction potential.
The Hamiltonian, with the multipole expansion of the potential
restricted to quadrupole terms, may be written, ~e.g.
\citet{hal90}, as follows:
\begin{equation}
    H(t)=H_0-\mathbf{d}\cdot \mathbf{F}(t)-1/6\sum_{jk}Q_{jk}F_{jk}(t) + \cdots,
\label{eq4}
\end{equation}
where the second term in Eq.~\ref{eq4}  describes the plasma-emitter
dipole interaction and the third represents the quadrupole
interaction of the radiator with the inhomogeneous
electric microfield of the plasma. The electric field $\mathbf{F}(t)$ and
the inhomogeneous microfield tensor $F_{jk}(t)$ represent the respective
total fields originating from ions and electrons.

We computed a pure H model atmosphere of typical WD parameters: $T_{\rm
eff}=12\;000 $ K (in figures below we use the abbreviation 12 kK) and $\log g=8.0$ (cgs units) 
using the ATM computer code.
Synthetic Balmer line
profiles were computed using our LINE code, which was derived from the
ATM program. Both codes solve the equation of state
in gas, assuming local thermodynamic equilibrium (LTE) and were described by \citet{mad83}.
They are practically the same as in papers by \cite{gra87}, hereafter ApJ'87,
and \citet{mad90}, hereafter A\&A'90, but were modified for the purposes of the present paper.
Numerical accuracy of the ATM code is high, since it ensures that
error of the bolometric radiation flux is much less than 1\% across the WD
model atmosphere. Both high accuracy and the physical foundations of these codes
(requirement of hydrostatic and radiative equilibrium) are comparable with
other model atmosphere computer programs used routinely nowadays, 
e.g. the computer code TLUSTY Version 195 \citep{hub88,hub97}  in its basic component, modelling plane-parallel, horizontally homogeneous stellar atmosphere in radiative and hydrostatic  equilibrium. LINE, which is a program for calculating the spectrum emergent from ATM model atmosphere, is comparable with program SYNSPEC Version 43
\citep{hub00}. The only serious simplification
is the LTE solution of the equation of state. Therefore, 
both ATM and LINE codes are suitable for differential investigations of 
the features of the synthetic Balmer-line profiles, when our interest is
focused on testing of two Stark opacity functions, distinctly differing 
from each other.    

Our model atmosphere of the sample white dwarf consists of 49 gas layers of 
various optical and geometrical depth. The coolest (shallowest) layer has
temperature $T = 5\; 500\; \hbox{K}$ and electron concentration about 
$N_e = 1.5 \times 10^{10}$ cm$^{-3}$, whereas the deepest (hottest) one,
contributing to the synthetic hydrogen spectral line profile, has
$T = 55\; 000\; \hbox{K}$ and  $N_e = 2 \times 10^{18}$ cm$^{-3}$. For each
layer of the WD model atmosphere we calculated complete Stark line opacity
profiles, which were broadened, {asymmetrical,} and shifted as the result of
all relevant physical processes in plasma, including motion and shielding
of perturbing ions plus the Stark-Doppler coupling (not only convolution, but also dependence). 
For this purpose we
applied the modified FCSM.

\section{Verifying the theoretical calculations}

In order to check the calculations of the complete --- broadened, asymmetrical, and shifted --- Stark profiles of the Balmer-$\alpha$ and Balmer-$\beta$ lines in {the proper} range of the physical conditions of fixed, homogeneous, thin-layer plasmas, we have carried out verifying experimental measurements. For the purpose of this study we have applied the same experimental device of the plasma generation (a high-current wall-stabilized arc), the light treatment, and the smoothing procedure of the spectral recordings, as those described in earlier papers: \citet{wuj02}, and \citet{gri05}. Specific data of the experiments were nearly the same as in the paper by \citet{gri05} --- with differences concerning the working gas (argon-hydrogen mixture in the cited reference) and the method of the plasma diagnostics (LTE), as follows: (1) As a plasma material we have used nearly pure argon, with a small admixture of the atomic hydrogen gas. (2) The {pLTE (partial LTE) plasma diagnostics} have been applied.

In the cylindrical channel of the discharge arc, working at the atmospheric pressure, the argon plasma is very stable, of nearly perfect cylindrical symmetry, and homogeneous along the lines of view parallel to the arc axis.  The argon plasma, with a small admixture of hydrogen, acquires features of high-density plasma, and, at the same time, the plasma is an optically thin layer in hydrogen lines. $T$ and $N_e$ values are extremely high along the symmetry axis of the arc channel, and decrease with the radial distance from the axis.
In the paper \citet{gri05}, we {diagnosed plasma conditions}
assuming local thermodynamic equilibrium (LTE). In the present paper, however, we carried out --- more careful and realistic --- {pLTE plasma diagnostics}, i.e. 
we adopted: modified Boltzmann's and Saha-Eggert's laws, Dalton's law and the electrical neutrality of the plasma. We took the integral emission coefficients of ArI 4300~\AA$\;$  and H$_{\alpha}$ or H$_{\beta}$  from the experiment. We have taken the overpopulation factors of ArI ground level for pLTE diagnostics  from the paper by \citet{hel81}.
The experimental investigations have been performed at plasma electron densities between $2.8$ and $10.0 \times 10^{16}$ cm$^{-3}$. Some results of the experiments are visible in Figs. 3 -- 6, as an example. In the matter of pressure shift and asymmetry of the investigated Balmer lines, we obtained an excellent agreement of the experimental and the calculated data.

\begin{figure}
\plotone{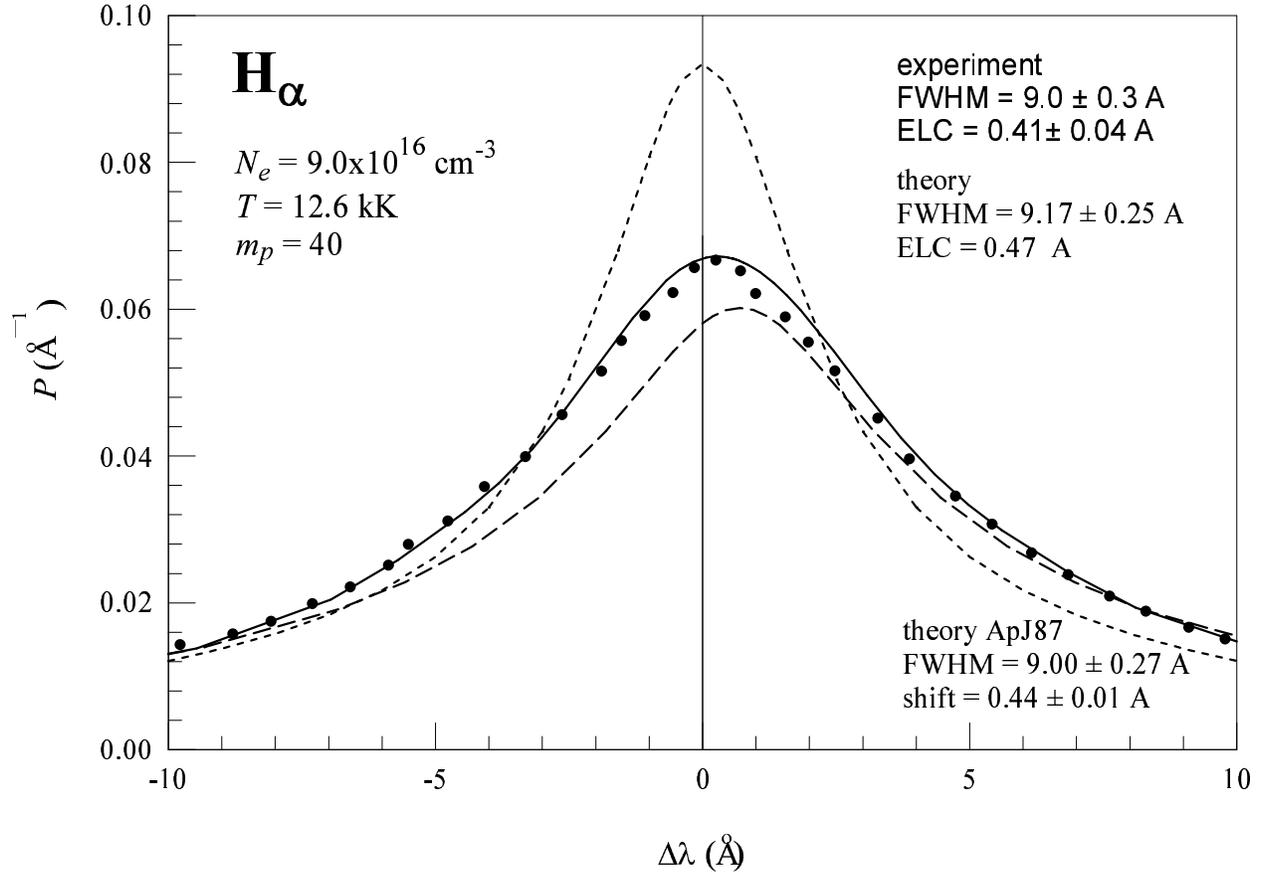}
\caption{Comparison of the H$_{\alpha}$ line profiles of different references and meanings in {the given} physical conditions of plasma ($m_p=40$ --- the atomic mass of argon). For details, see text.\label{fig3}}
\end{figure}

In Fig.~\ref{fig3}, the Stark line-profiles of the H$_{\alpha}$ are compared. The black points --- the experimental data of the present paper; the long-dashed curve --- the same theoretical approximation as that used in our paper ApJ'87 --- symmetrical Stark profile  following \citet{edm67}, hereafter  ESW, but shifted as a whole,
in dependence on the physical conditions of plasma in the given layer of the WD atmosphere; the short-dashed curve --- Stark line-profile by \citet{vid73}, hereafter VCS --- a symmetrical and unshifted one; this approximation is the main one in the astrophysical spectroscopy (e.g. \citet{bar05}); the solid curve --- our modified-FCSM, complete Stark line-profile of Balmer H$_{\alpha}$, asymmetrical and shifted to the red. We see particularly that: {(1)} Agreement of our calculated (using mFCSM)
 and measured Stark line-profile of Balmer H$_{\alpha}$ 
is a perfect one; {(2)} ESW ---  shifted (ApJ'87) --- the line profile runs correctly; {(3)} VCS line-profile has a defectively described line center, and is not suitable for WD applications.  

{The FWHM of the line} and the red shift of the ``experimental line center'' \citep[ELC, as defined by][]{wie71} of the experimental line-profile and of the modified-FCSM one are also shown in the figure. The agreement of these data in the case of the H$_{\alpha}$ line, as well as the H$_{\beta}$ line (Fig.~\ref{fig4} below), is excellent. We would like to pay our attention to the pressure shift of the H$_{\alpha}$ line peak --- a little smaller one in the experimental and in modified-FCSM line-profiles than that in paper ApJ'87, similarly as in the case of the H$_{\beta}$ line, discussed in detail in Figs.~\ref{fig5} and \ref{fig6} below. 
 
In Fig.~\ref{fig5}, we see that in ApJ'87 the PS-values of the peak of the H$_{\alpha}$ Stark line-profile are considerably overestimated. In the case of {the H$_{\beta}$} line, the circumstances are different. In Fig.~\ref{fig6}, it is clearly seen that PS of the central dip of H$_{\beta}$ Stark line-profiles were predicted 
well in ApJ'87, with a little excess as compared with the present, more precise data. The PS measured by \citet{wie72},  presented in Fig.~\ref{fig6}, require a comment. We see that their best-fitting curve achieves zero electron concentration (i.e. the conditions of the isolated atom) at the blue pressure shift of about $0.2$~\AA, which is, of course, not a reasonable result. Probably all these measurements of the H$_{\beta}$ line were taken using temporarily defected recording equipment, producing unphysical, systematic (additive) red shift. (We see that Wiese's and our curves of the PS are nearly parallel.) When the presumed additive contribution (probably erroneous) is removed, the measurements by Wiese at al. precisely agree with our measurements and the modified-FCSM calculations.

\section{The emerging (synthetic) WD line-profile calculations}

Complete --- broadened, {asymmetrical,} and shifted --- Stark line-profiles have been calculated for each layer of the WD atmosphere under examination, and, subsequently, convolved with the proper Doppler contribution. In Figs \ref{fig7} and \ref{fig8}, we see, as an example, the final convolution of Stark-Doppler effect for one of the shallowest layers (a relatively cool and rare one) --- Fig.~\ref{fig7}, and for one of the deepest  (hot and dense) atmospheric layers --- Fig.~\ref{fig8}.

The emerging (synthetic) Balmer line-profile is a weighted mean, balanced by the transfer equation, of the contributions of all active, outer layers of the pure H model atmosphere. The superposition of the complete Stark-Doppler profiles of the H$_{\beta}$ Balmer line (asymmetrical and red shifted), corresponding to successive atmospheric layers --- from the extremely outer layer (narrow structure)  down to the deepest one (extremely broad shapes) --- is shown in Figs.~\ref{fig9} and \ref{fig10}. (The H-alpha line was treated just in the same maner.)

In Fig.~\ref{fig11}a, the synthetic spectrum in the region of {the H$_{\beta}$ line}, calculated using opacity profiles, as in Figs.~\ref{fig9} and \ref{fig10}, is presented in the fluxes scale. This ``wide-angle'' spectroscopic view is, of course, very similar to those observed via most advanced present-day instrumentation, as reported, e.g., by \citet{bar01}, Fig.~\ref{fig4}, ground-based; \citet{bar05}, Fig.~5, HST; or \citet{kep07}, Figs.~10 and~11, SDSS. In Fig.~\ref{fig11}b --- the same, but in a reduced (0-1) scale, 
$r_{\lambda}=\hbox{Flux}_{\hbox{\scriptsize line}}/\hbox{Flux}_{\hbox{\scriptsize continuum}}$. 
From the point of view of the gravitational red shift measurements, the clue lies in the line center ---  in the narrow central section, barely a few {\AA}ngstr{\"o}ms  wide. We discuss the final results of the analysis of that region  in detail in the next section.

\section{Results and conclusions}

The pressure shift in the central parts of the synthetic H$_{\alpha}$ Balmer line is practically negligible; our earlier (ApJ'87) diagnosis in that mater is hereby confirmed. \citep[N.B.: in very-high-density plasma conditions that scenery is more complicated --- cf. the paper by][e.g. Figs.~9 and~13 therein --- the pressure shift of the H$_{\alpha}$ line achieves substantial values.]{gri05} In the case of the synthetic H$_{\beta}$ line, the central region is slightly affected by the pressure shift, as correctly predicted in ApJ'87, but the near and far wings are affected enormously, as compared with ApJ'87. It is seen in our Figs. \ref{fig12} and \ref{fig13},
plotted on the basis of the synthetic spectra of the WD atmospheres of the typical parameters: $\log g = 8.0$, and $T_{\rm eff}=10\;000 $ K (in figures we have used abbreviation: 10 kK), as an example.

We consequently use the concept of the pressure red shift of the WD Balmer lines as the red shift of the ``averaged line center'' (ALC), introduced by \citet{gre67} to the WD spectra, and adapted in the experimental measurements, as the ELC paradigm, by \citet{wie71} and \citet{wie72}. So, in Figs.~\ref{fig12} and \ref{fig13} on the abscissa axis the quantity $\Delta \lambda$~(\AA) is used; on the ordinate axis --- the pressure red shifts in the equivalent Doppler velocity units ($v_F$ or $v_r$, respectively), in conformity with the tradition in that subject. In Fig.~\ref{fig12}, the lower index $F$ at the velocity symbol reminds that the ordinate axis of the synthetic WD spectrum is the fluxes scale $F_{\lambda}$ (not the residual intensity $r_{\lambda}$). The thick curve in Fig.~\ref{fig12} has been constructed by us from the WD synthetic spectrum, strictly in the same manner as an observer does on raw recording of the real WD spectrum. Here we must note, however, that the straightforward measurement of the WD gravitational red shift in ALC-convention can be falsified by apparent, ``trivial'' causes, the principal of which is the red shift induced by the sloping of the Paschen continuum \citep[cf., e.g.,][]{sch77}. The measurements  in the residual intensities, $r_{\lambda}$, in the (0-1) scale, are free of that ``trivial'' contribution.

In Fig.~\ref{fig13}, the main results of this paper --- the Stark-induced red shift of the investigated Balmer lines {for a WD} of {the given} parameters --- in the reduced scale, r$_{\lambda}$, are presented. We see here four curves. One pair of the curves correspond to H$_{\alpha}$, another --- to {the H$_{\beta}$ line}. The solid lines show new effects --- the results of the present paper; the dashed ones refer to the results of ApJ'87, with the aim of comparing these results. Each of the shown dependences presents the Stark-induced red shift (pressure shift, PS) alone, i.e. net pressure line shift (free of ``trivial'' contributions), produced by physical, elementary processes in plasma of the WD atmosphere, in all its layers active in the formation --- \textit{via} the transfer equation --- of the synthetic shape of a Balmer line. 

We remember here the fundamental differences in the approaches applied in the earlier papers ApJ'87, A\&A'90 and in the present paper: In the former ones, the symmetrical line-profiles of the Balmer lines, shifted as a whole, were used. The pressure shift of the line center, used at that time, remains nearly correct. However, the synthetic line-profiles --- calculated at that time using symmetrical line-profiles, shifted as a whole --- have been uncertain.  In the present paper, the Stark-line profiles used are not only red shifted but also asymmetrical ones, similarly as in real conditions. The results presented in Fig.~\ref{fig13}  permit us to formulate the following conclusions: {(1)} The Stark-induced red shift (pressure shift, PS) of the H$_{\beta}$ line in WD spectra must be taken into account when the gravitational red shift measuring, especially in cases when the ALC paradigm is applied. In the H$_{\beta}$ line in WD spectra, the Stark-induced red shift is far greater than that in the case of {the H$_{\alpha}$ line}; in the case of the last one, the pressure red shift can practically be neglected. {(2)} Our new, values of the ``observed'' pressure red shift of the WD H$_{\alpha}$ line are nearly twice smaller than those predicted in ApJ'87, whereas in the case of {the WD} H$_{\beta}$ line --- considerably greater, especially at $\Delta\lambda > 10$~\AA, compared with those in ApJ'87. 

\begin{figure}
\plotone{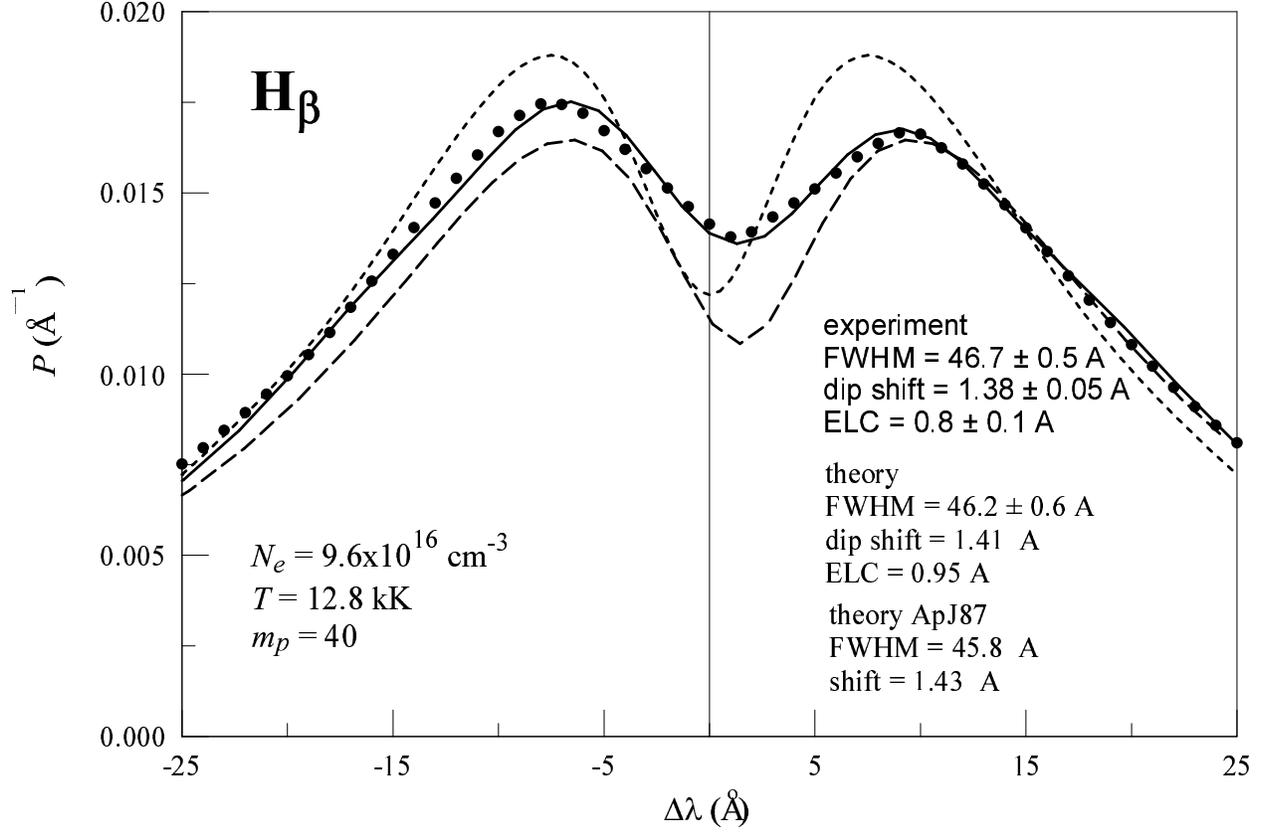}
\caption{Comparison of the H$_{\beta}$ line profiles of different references and meanings in {the given} physical conditions of plasma. The symbols and the curves  used have the same sense as in Fig.~\ref{fig3}. 
The central {dip} displayed on the line-profiles represented by the dashed curves is  too deep. This defect is the consequence of the negligence of the ion dynamics.  The agreement of the measured and the calculated mFCSM data is excellent. \label{fig4}}
\end{figure}

The effects of such clear differences in the observed features of both these lines, and in the resulting Stark-induced red shift in WD spectra, can be explained as a result of dissimilarity of the quantum structures of both lines. The H$_{\alpha}$ line belongs to the so-called ``odd'' Balmer lines, with a strong central line-component, whereas {the H$_{\beta}$ line} --- to ``even'' ones, without the central component. These differences strongly manifest themselves in the laboratory (emissive, and thin-layer plasma; cf., e.g., Figs. \ref{fig3} and \ref{fig4}) plasma spectroscopy. In the infinitely thick stellar atmospheres, the ``odd-even'' problem is --- from the point of view of the observer --- completely masked by processes of the transfer of radiation to outer layers of the stellar atmosphere and to the observer's eye: the synthetic lines of both these kinds have similar shapes, with a clearly formed central dip. In each atmospheric layer, however, the line-shapes of opacities have such features as in a thin layer --- cf. Figs.~7--10 of the present paper. 

The H$_{\beta}$ line-shape is a specific one (see Figs.~\ref{fig4} and~\ref{fig8}) --- the red peak of the line is lower than the blue one (i.e., the ``blue'' opacity is greater than that in the ``red'' line region). Consequently, a radiation transfer in the region of red peak proceeds more easily compared with that in the blue peak. In the red side of the line, we ``see'' therefore more deep, dense, and hot layers; on the blue side of the line --- more shallow, relatively rare and cool layers. These effects averaged (by a radiative transfer equation) produce additional (as compared with contribution of a single atmospheric layer) asymmetry and Stark-induced red shift of the synthetic line-profile. The case of {the H$_{\alpha}$ line} is, due to a strong central component and nearly symmetrical Stark splitting, incomparably simpler; asymmetry effects (appearing in this line too), do not produce any substantial Stark-induced red shift.     

When spectroscopic data of extremely high dispersion are available, e.g. from HST or VLT, and when {the H$_{\beta}$ line} with the central non-LTE dip (cf., e.g., Falcon et al. 2010, Fig.~2, or Casewell et al. 2009, Figs.~2 and~3) measurement is used, the problem of the Stark-induced shift can be completely ignored. The non-LTE dip is formed in the uppermost, extremely rare atmospheric layers of WD, where Stark effects are of no importance as against Doppler effect (cf. Fig.~\ref{fig7}). Here the observed line-shift is entirely caused by the gravitational (Einstein's) effect. 
In more frequent cases of the middle-resolution spectra, for example from SDSS, \citep[e.g.,][]{kle04, kep07, mad04},  at {the gravitational} red shift measurements the ``average line center'' defined in broader regions of {the H$_{\beta}$ line} should be used. The same concerns the older archival spectroscopic data, often invaluable records of  the past days. 

When the synthetic H$_{\alpha}$ or H$_{\beta}$ Balmer lines-profiles in the residual intensity units, $r_{\lambda}$, in {the (0-1)} scale are used (as we see, e.g., in Fig.~\ref{fig11}b), the data from Fig.~\ref{fig13}, free of all ``trivial'' agents and most reliable ones, should be used throughout.

We hope that the presented results will be useful in a more precise determination of the gravitational red shifts in the WD spectra and in a more accurate determination of the WD mass/radius ratio. It is important from the point of view of our understanding of WDs --- objects
 largely widespread in Nature. Better knowledge of that ratio can also be helpful in imposing new limitations on the WD interior theory and can faciliate better calibration of Ia type supernovas --- exploding WDs in binary systems --- and, consequently, can permit to more precisely examine the density of the dark matter in the Universe.

\section{Acknowledgements}

JM acknowledges support by Polish National Science 
Centre grants No. 2011/03/B/ST9/03281, and   2013/10/M/ST9/00729.

\clearpage

\clearpage

\clearpage

\clearpage

\clearpage

\begin{figure}
\plotone{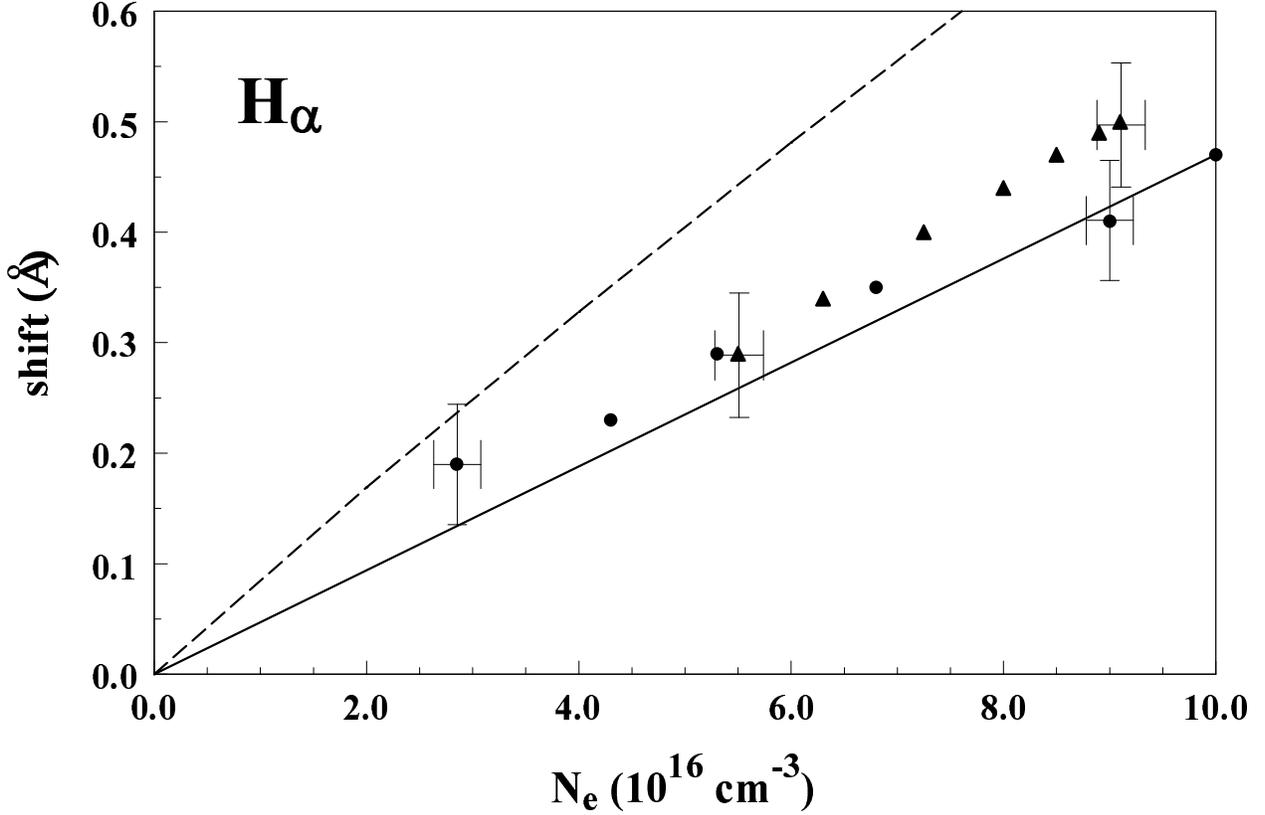}
\caption{The pressure shift (PS) to the red of the peak of the H$_{\alpha}$ line as a function of the electron concentration in plasma. The dashed curve --- the PS as in ApJ'87; the solid curve --- PS of our present calculations using mFCSM; the points: the black triangles --- measurements by \citet{wie72}, the full circles --- our measurements, both according to the paradigm of ELC.\label{fig5}}
\end{figure}

\clearpage

\begin{figure}
\plotone{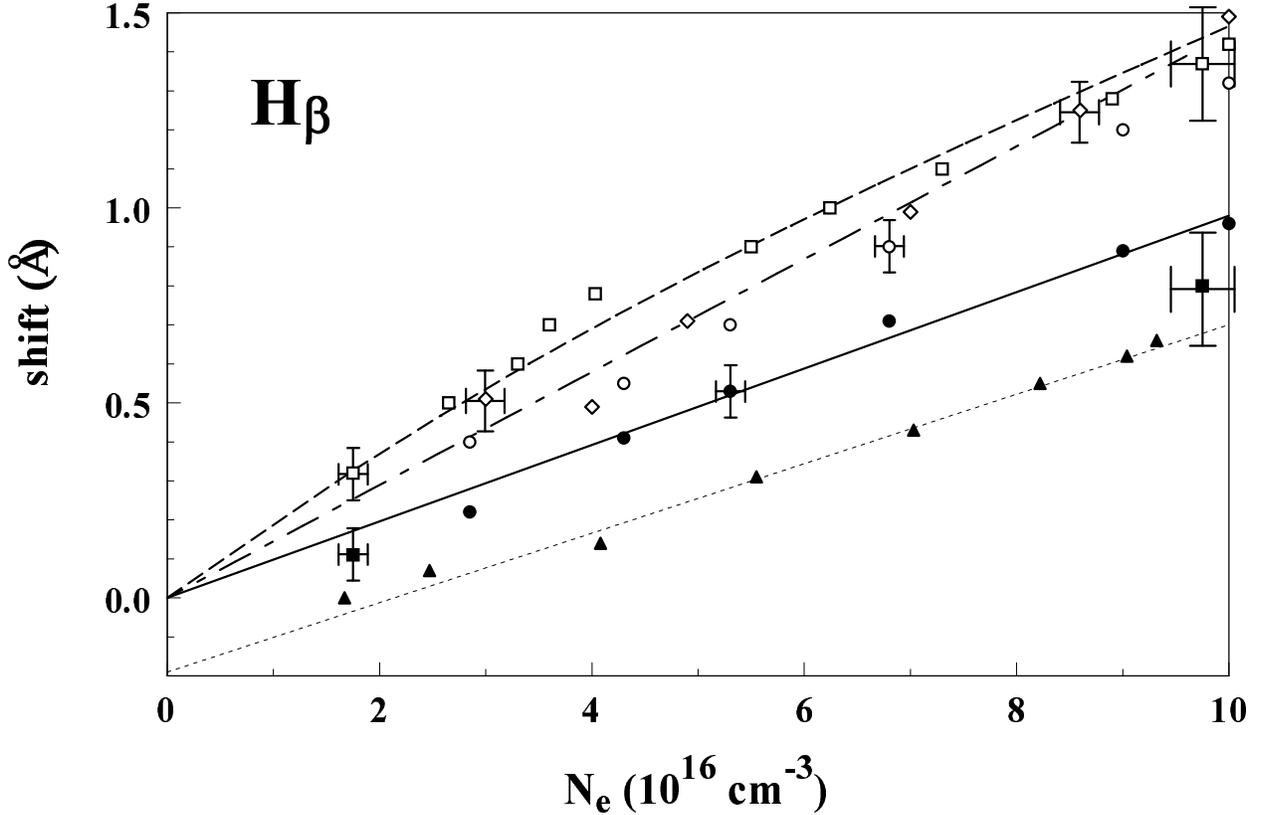}
\caption{The pressure shift (PS) to the red of the central dip of the H$_{\beta}$ line as a function of the electron concentration in plasma. The lines and the symbols have the following meanings: the black triangles and the dotted line --- the original experimental measurements of the ELC and the proper best fit, respectively, by  \citet{wie72}; the solid line and full circles --- the ELC-type red shifts, calculated (via mFCSM) and measured,  respectively, in the present paper; the short-long-dashed curve --- the pressure red shift of the central dip according to the mFCSM calculations in the present paper; the short-dashed curve --- the PS of the central dip as in ApJ'87. The points, the shapes of which are not commented on here, are taken from different bibliographical positions and, as a rule, belong to foreign authors.\label{fig6}}
\end{figure}

\clearpage

\begin{figure}
\plotone{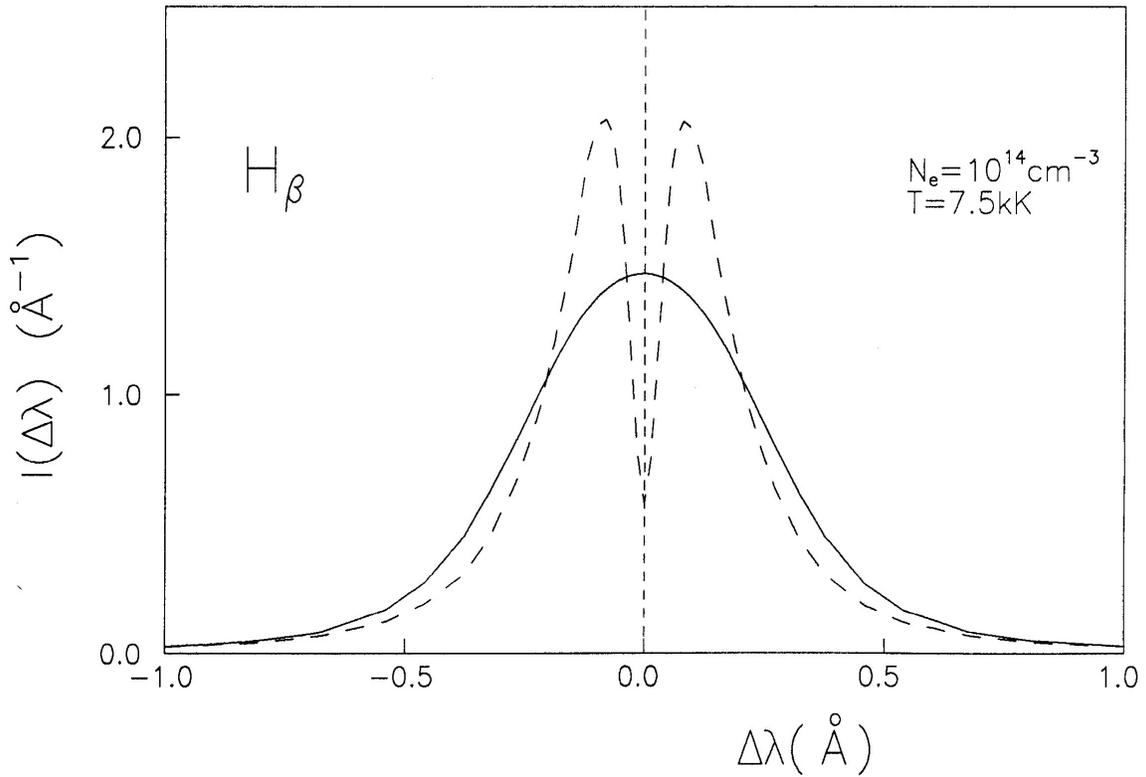}
\caption{The Stark H$_{\beta}$ shape (the dashed line) and the final effect of the Stark-Doppler
 convolution (the solid line) in one of the shallowest WD atmospheric layers. We see that the final shape, given by the convolution operation, differs completely  from the initial Stark shape. This is of the Doppler-type one, without any central structure.\label{fig7}}
\end{figure}

\clearpage

\begin{figure}
\plotone{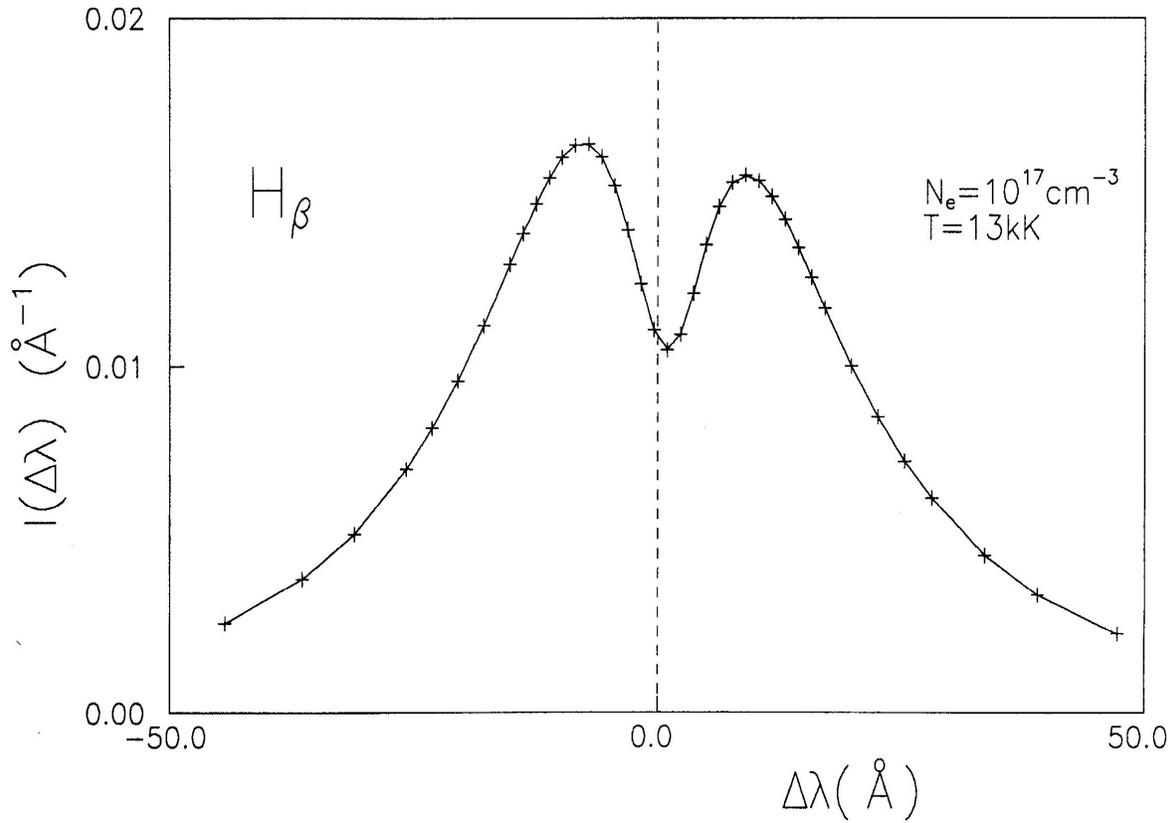}
\caption{The Stark H$_{\beta}$ shape (the solid line) and the final effect of the Stark-Doppler convolution (crosses) in one of the deepest WD atmospheric layers. We see that the shape, given by the convolution operation, is identical with the initial Stark shape. In deep, hot and dense atmospheric layers, where Stark broadening dominates totally, the contribution of the Doppler  effect is completely negligible.\label{fig8}}
\end{figure}

\clearpage

\begin{figure}
\plotone{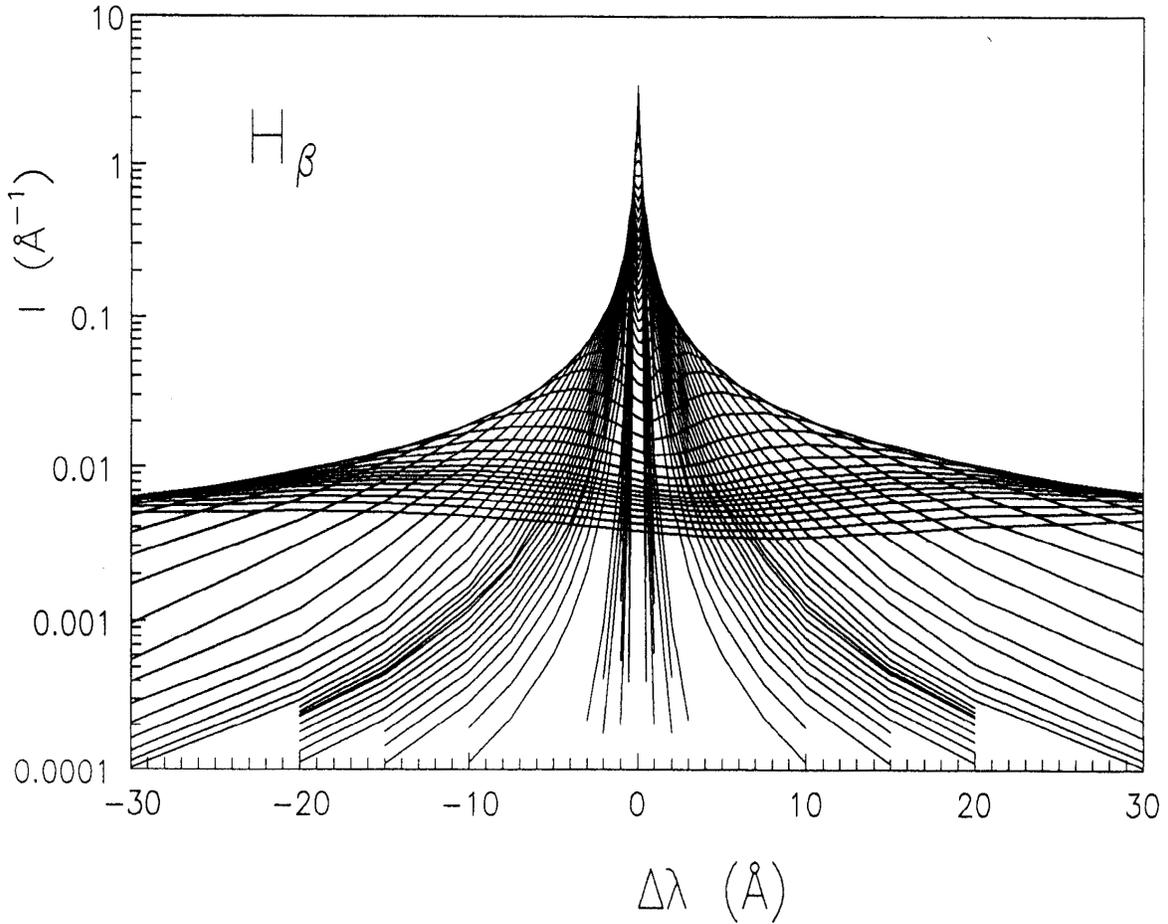}
\caption{Superposition of the convolved Stark-Doppler profiles, corresponding to the successive WD atmospheric layers --- from the outer most  layers (the narrow, symmetrical, and unshifted shapes), down to the deepest ones (the extremely broad, asymmetrical, and red shifted shapes). The line-shape evolution is monstrous, indeed.\label{fig9}}
\end{figure}

\clearpage

\begin{figure}
\plotone{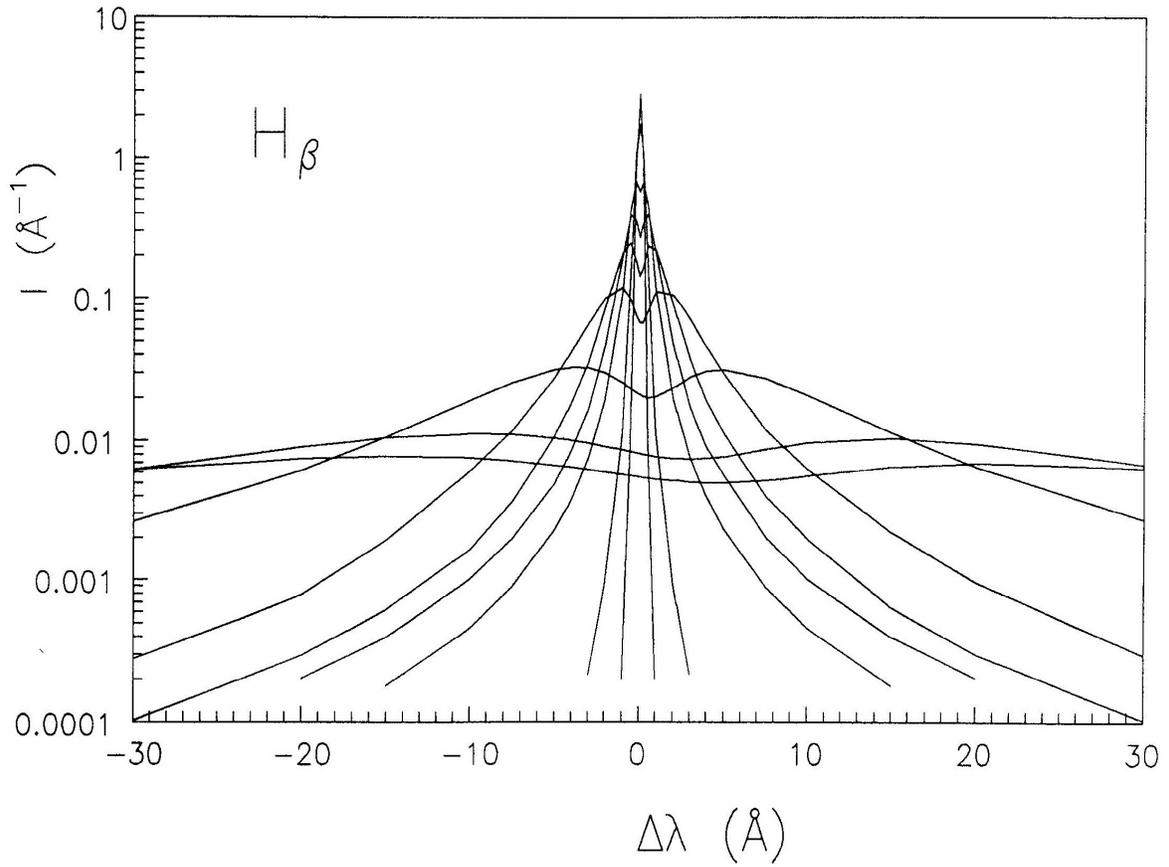}
\caption{The same superposition as in Fig.~\ref{fig9}, but, for clarity, 
containing only a few line profiles (H$_{\beta}$  opacity shapes)
from selected --- every fifth --- atmospheric layers.
\label{fig10}}
\end{figure}

\clearpage

\begin{figure}
\plottwo{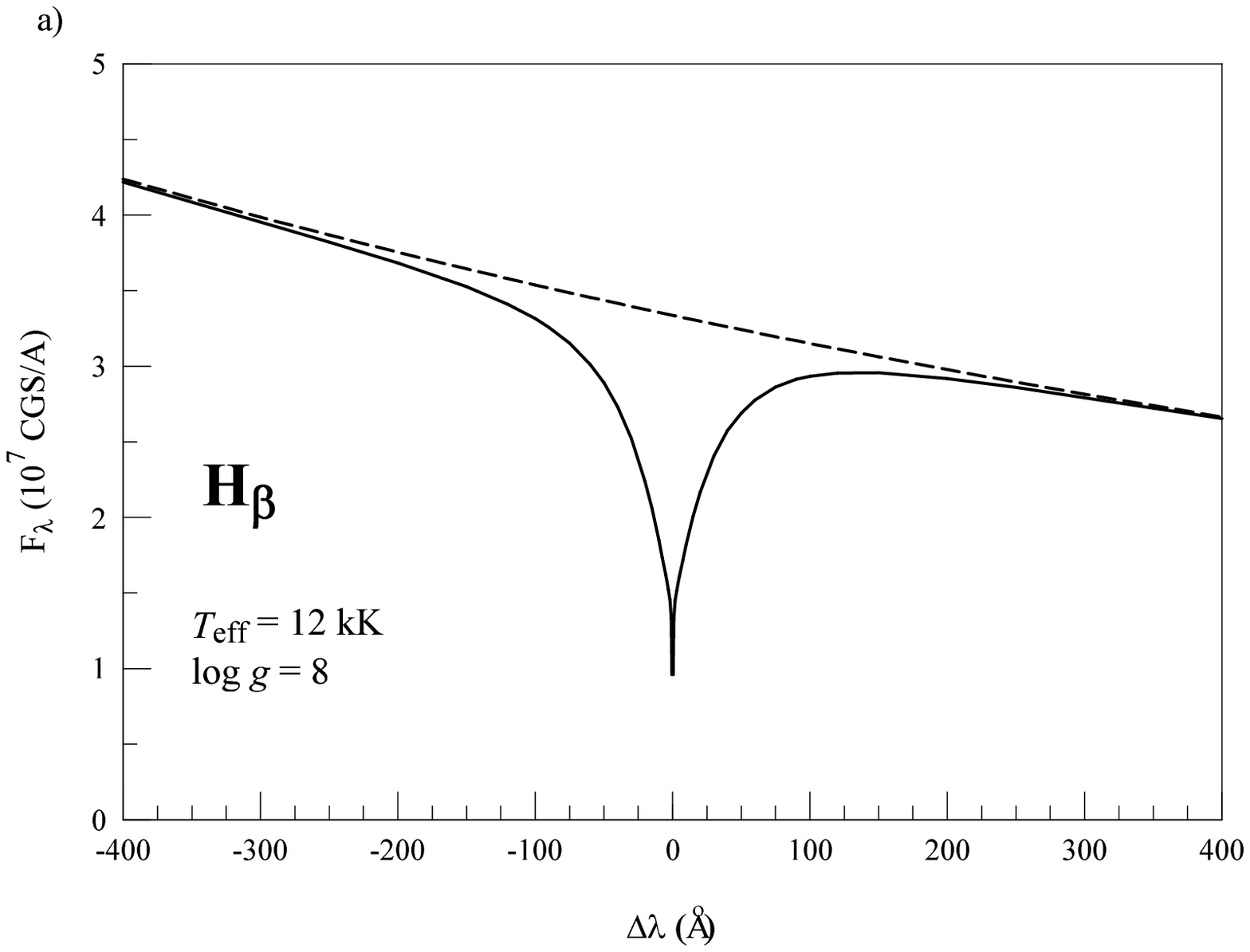}{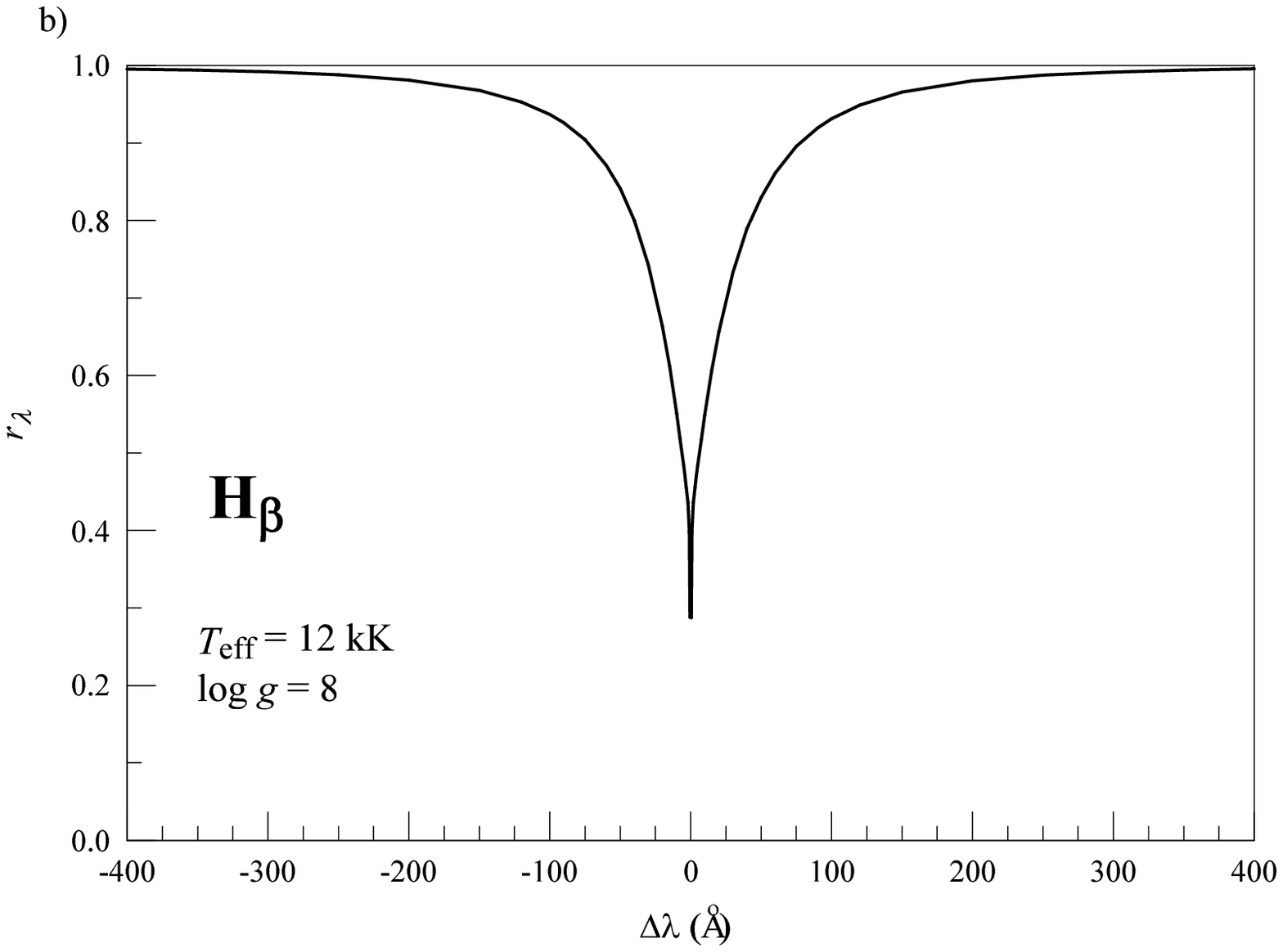}
\caption{The synthetic  spectrum in the region of the H$_{\beta}$ line of a virtual WD of the given parameters, in the fluxes scale (left) and in the reduced (0-1) scale (right).\label{fig11}}
\end{figure}

\clearpage

\begin{figure}
\plotone{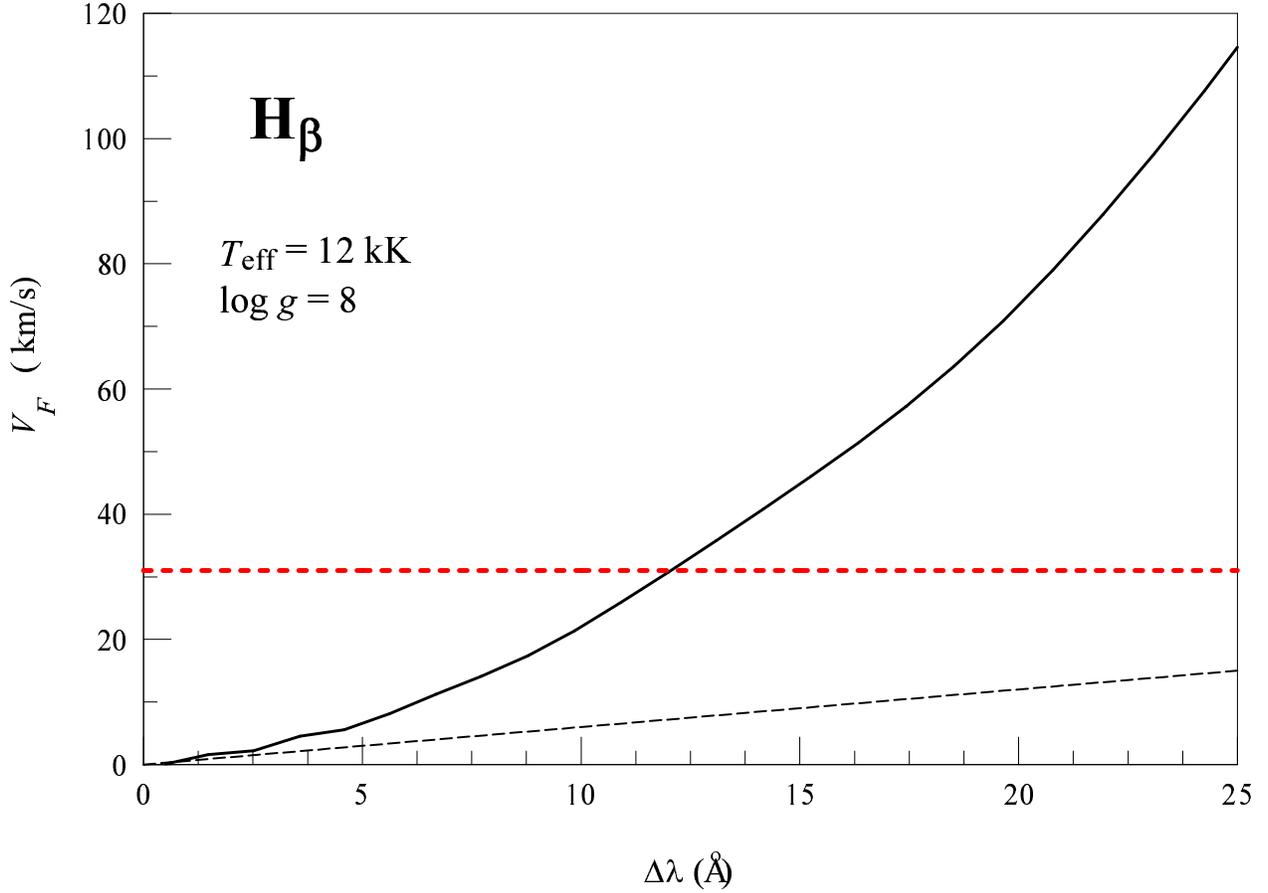}
\caption{The Stark-induced shift (the pressure shift, PS) of the Balmer H$_{\beta}$ line, in ALC paradigm, of the WD atmosphere of {the given} parameters, as a function of the distance from the line center, in the fluxes scale.
This figure has been generated on the basis of the diagram of the type of Fig.~\ref{fig11}a. The solid line shows the present result, the thin dashed line --- that in ApJ'87. For details, see the text. The dashed (horizontal) line represents approximately the level of the gravitational red shift in WDs of the parameters, as in the legend above --- cf., e.g. \citet{pro98}, \citet{fal10}.
\label{fig12}}
\end{figure}

\clearpage

\begin{figure}
\plotone{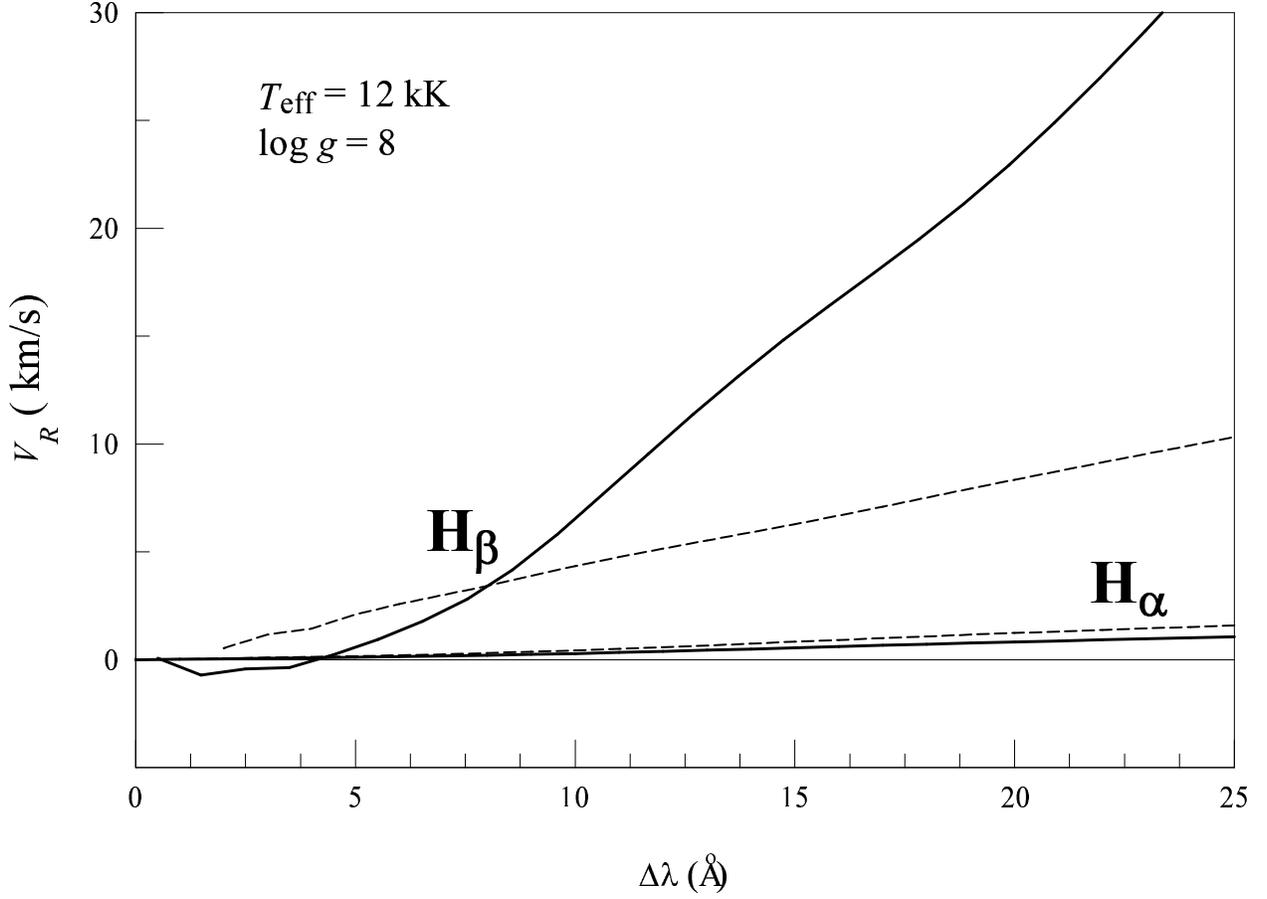}
\caption{The main results of the present paper --- the Stark-induced red shifts, 
expressed in the Doppler velocity units, as functions of the distance from the line center, in the synthetic spectra of H$_{\alpha}$  and H$_{\beta}$  Balmer lines of the WD of {the given} parameters.
The diagram has been plotted on the basis of the synthetic Balmer line-profiles, expressed in
the residual intensities, $r_{\lambda}$, in (0-1) scale of the type of Fig.~\ref{fig11}b. For detail, see the text.\label{fig13}}
\end{figure}

\end{document}